# Tirocinio Formativo Attivo: il Laboratorio Pedagogico Didattico TIC della Classe A049 nell'Ateneo di Palermo

**Lucia Lupo[1], Aurelio Agliolo Gallitto[2]**
[1]*Liceo Scientifico Statale "Galileo Galilei", via Danimarca 54, I-90146 Palermo, Italy*
[2]*Dipartimento di Fisica e Chimica, Università di Palermo, via Archirafi 36, I-90123 Palermo, Italy*

**Abstract**
We discuss the organization of the course of Laboratorio Pedagogico Didattico TIC of the Tirocinio Formativo Attivo at the University of Palermo. The course takes into account the development of information and communication technologies both for the management/administration of school and their application in Mathematics and Physics education. In particular, we will discuss in detail the various possibilities of renewing the teaching given by communication and sharing tools of web 2.0, the use of the digital whiteboard, and the use of technologies in Mathematics and Physics laboratories, up to the use of multimedia mobile devices as experimental tools.

**Riassunto**
Con il secondo ciclo di Tirocinio Formativo Attivo è stato attivato l'insegnamento di Laboratorio Pedagogico Didattico TIC. Nell'articolo discuteremo come è stato organizzato questo insegnamento al fine di poter affrontare problematiche legate all'uso delle tecnologie dell'informazione e della comunicazione sia nella gestione amministrativa della Scuola sia in ambito didattico. Discuteremo in particolare le varie possibilità di rinnovamento della didattica introdotte dagli strumenti di comunicazione e condivisione del *web* 2.0, dall'uso della lavagna interattiva multimediale, fino all'uso dei dispositivi portatili multimediali come strumenti di laboratorio.

# 1. Introduzione

Con il D.M. 487 del 20/06/2014 è stato attivato nell'A.A. 2014/2015 il secondo ciclo di Tirocinio Formativo Attivo (TFA) di abilitazione alla professione di docente. Una novità del D.M. 487, rispetto al primo ciclo, è stata l'introduzione di un laboratorio dedicato alle tecnologie dell'informazione e della comunicazione (Laboratorio Pedagogico Didattico TIC) e di un laboratorio dedicato al settore dei bisogni educativi speciali (Laboratorio Pedagogico Didattico BES) da svolgersi nell'area specifica, per almeno 15 ore di attività, separatamente dagli altri insegnamenti, lasciando invariata la distribuzione del carico didattico sia nell'area delle didattiche disciplinari sia nell'area comune [1]. Nell'Ateneo di Palermo, il Comitato di Coordinamento ha deciso di riproporre la stessa offerta formativa del TFA I Ciclo [2] e dedicare 1 Credito Formativo Universitario (CFU), della durata di 15 ore, sia al Laboratorio TIC sia al Laboratorio BES. Questi 2 CFU sono stati recuperati riducendo di 1 CFU l'insegnamento di Storia della Fisica e l'insegnamento di Storia della Matematica.

Il Laboratorio TIC è stato organizzato in una prima parte dedicata alle novità nella gestione della Scuola e in una seconda parte dedicata alle tecnologie dell'informazione e della comunicazione nella didattica. Sono state trattate le possibilità di rinnovamento della didattica introdotte dagli strumenti di comunicazione e condivisione del *web* 2.0, dall'uso della lavagna interattiva multimediale (LIM), dalla tecnologia nel laboratorio di Matematica e nel laboratorio di Fisica, ponendo l'attenzione alla più recente sfida della didattica tecnologica rappresentata dall'uso dei dispositivi portatili multimediali (DPM) come strumenti di laboratorio. In questo articolo descriveremo le principali attività svolte nell'ambito del Laboratorio TIC del Corso di TFA della Classe A049 Matematica e Fisica, nell'Ateneo di Palermo.

# 2. Le tecnologie per l'informazione e la comunicazione in ambito scolastico

Il mondo della Scuola è da qualche anno interessato a un processo di "informatizzazione" che riguarda sia l'organizzazione amministrativa sia la comunicazione fra cittadino e Istituzioni. Con l'intento di operare uno snellimento nei procedimenti amministrativi, il MIUR ha attivato il sistema di presentazione online delle istanze, Progetto POLIS [3]; per rendere trasparente e accessibile il patrimonio dati ha costituito la banca dati "Scuola in chiaro" [4]. A livello locale, ogni Scuola è tenuta a pubblicare nell'albo pretorio online notizie e avvisi di interesse pubblico. In seno al processo di "dematerializzazione" [5], che punta a limitare sempre più l'uso della documentazione cartacea nella Pubblica Amministrazione, il MIUR ha reso operativo il portale dei servizi SIDI [6], attraverso il quale le famiglie provvedono all'iscrizione degli alunni e le segreterie alla gestione dei dati sulle iscrizioni; ogni Istituto Scolastico si è dotato di *software* per la gestione dei registri *online*, degli scrutini elettronici, della comunicazione alle famiglie tramite *web* e posta elettronica.

Già da tempo la Scuola comunica con il territorio attraverso i siti *web*; negli ultimi anni, le potenzialità espresse dalle applicazioni sviluppate nel *web* 2.0 hanno permesso l'ampliamento di questa comunicazione tramite *blog* e *forum*. Diversi Istituti Scolastici hanno un proprio profilo sui *social network*, spesso gestito dagli stessi alunni e molte hanno anche reso disponibili piattaforme per l'*e-learning*. La possibilità di usare ambienti di formazione a distanza (FAD) fa sì che la Scuola possa operare in casi in cui gli alunni siano costretti a prolungate assenze dovute a ospedalizzazioni o ad altre esigenze personali rientranti nell'ambito dei Bisogni Educativi Speciali. Anche in assenza di specifiche esigenze, la Scuola può ampliare la propria offerta formativa con la creazione di classi virtuali nelle quali docenti, discenti e famiglie possono tenersi in contatto, scambiare materiali, realizzare percorsi alternativi rispetto alla prassi quotidiana avvicinandosi all'esperienza degli alunni di oggi, appartenenti alla cosiddetta "generazione dei nativi digitali" [7].

Nel Laboratorio TIC, i tirocinanti della Classe di Abilitazione A049, oltre a documentarsi sulle tecnologie dell'informazione e della comunicazione nell'ambito della gestione della Scuola, hanno sperimentato la piattaforma di condivisione EDMODO ideata proprio per la didattica [8], analizzandone le potenzialità e i limiti. I tirocinanti, anche se non rientrano nella generazione dei nativi digitali, hanno notevoli esperienze nel campo della comunicazione via *web* e la loro analisi della piattaforma è stata molto critica: hanno rilevato come *bug* la mancanza dell'aggiornamento automatico della pagina e la difficoltà nella condivisione di documenti, cosa che invece è risultata agevole con altri strumenti di condivisione via *web*, quali per esempio Google Drive e Dropbox.

**2.1. Lavagna interattiva multimediale e libri elettronici**

La lavagna interattiva multimediale (LIM) è di fatto uno schermo digitale *touchscreen* integrato in un sistema *hardware/software* costituito da un videoproiettore e un computer su cui è installato un *software* di gestione specifico. Alle funzioni del sistema classico computer/proiettore/schermo, la LIM aggiunge le potenzialità del *touchscreen* e la possibilità per il docente di creare oggetti didattici, di registrare i propri interventi in classe e di inviare in tempo reale agli alunni i materiali prodotti [9]. L'uso efficace del sistema LIM è connesso quindi alla capacità progettuale del docente e agli strumenti disponibili, quali per esempio il libro elettronico (*e-book*) introdotto ufficialmente nella Scuola con la Circolare N.18 del 9 febbraio 2012: "*Le adozioni da effettuare nel corrente anno scolastico, a valere per il 2012/2013, presentano una novità di assoluto rilievo, in quanto, come è noto, i libri di testo devono essere redatti in forma mista (parte cartacea e parte in formato digitale) ovvero debbono essere interamente scaricabili da internet. Pertanto, per l'anno scolastico*

*2012/2013 non possono più essere adottati né mantenuti in adozione testi scolastici esclusivamente cartacei*".

Il libro elettronico è uno strumento di lavoro che presenta molti vantaggi anche nel caso in cui resta una semplice trasposizione del libro in formato cartaceo: può essere letto su *tablet* e *smartphone*, riducendo il problema dell'ingombro e del peso e può essere utilizzato dal docente con la LIM. Spesso i progetti editoriali delle case editrici ampliano i contenuti del libro in forma cartacea con collegamenti a siti di interesse, ambienti di simulazione di laboratorio, di esercitazione e studio personalizzato.

Nel percorso formativo proposto ai tirocinanti, è stata analizzata da diverse prospettive la questione della seconda prova scritta dell'esame di Stato nel Liceo Scientifico, in quanto a partire dall'anno scolastico 2014/2015 la seconda prova scritta ha subito delle modifiche nella forma, non nella struttura, in accordo con quanto previsto dalla Riforma Gelmini, che in questo anno scolastico ha concluso la fase transitoria [10]. I tirocinanti hanno partecipato nelle scuole accoglienti alla simulazione proposta dal Ministero e, nell'ambito del Laboratorio TIC, hanno riflettuto sulle possibilità di arricchimento della preparazione degli studenti, analizzando a titolo esemplificativo la piattaforma MATutor predisposta dalla Zanichelli per la preparazione alla prova [11].

## 3. Sensori *real time* e dispositivi portatili multimediali

Nella seconda parte del Laboratorio TIC sono stati affrontati alcuni aspetti legati all'uso di sistemi tecnologici innovativi nelle metodologie laboratoriali per l'insegnamento/apprendimento della Matematica e della Fisica, con particolare attenzione all'innovazione introdotta dai dispositivi portatili multimediali (DPM).

### 3.1 Sensori *real time*

La diffusione dei computer ha dato la possibilità di attrezzare i laboratori didattici per eseguire una notevole varietà di esperimenti scientifici grazie all'uso di sensori connessi al computer per mezzo di una interfaccia *hardware*. Per esempio, l'interfaccia LabPro della Vernier consente di collegare al computer una notevole quantità di sensori, acquisire automaticamente i dati rilevati dai sensori ed elaborarli tramite il *software* LoggerPro, mentre il dispositivo palmare LabQuest può essere usato sia come interfaccia sia in modo autonomo per l'acquisizione di dati anche in situazioni *outdoor* [12]. In questo modo è possibile analizzare i dati raccolti in tempo reale e, per questo motivo, tali strumenti sono denominati sensori *real time*. L'evoluzione di queste tecnologie è rapida e si è passati in pochi anni da sensori che si connettono al PC tramite interfacce esterne, a sensori che si connettono direttamente al computer tramite porta USB (come i sensori Go!Motion e

Go!Temp della Vernier), all'ultima frontiera rappresentata dai sensori che si connettono in modalità *wireless* [13]. Allo sviluppo della tecnologia si è affiancato l'abbattimento dei costi, che ha negli ultimi anni permesso di rinnovare la strumentazione dei laboratori scolastici.

I tirocinanti hanno avuto modo di sperimentare l'uso di questa tecnologia negli insegnamenti disciplinari; nel Laboratorio Pedagogico Didattico TIC è stato presentato, a titolo esemplificativo, l'uso del sensore di posizione Go!Motion come attività introduttiva per la costruzione del concetto di posizione e velocità nello studio della legge oraria nel moto rettilineo. Questa attività è stata scelta per la sua facile realizzazione, anche in una semplice aula dotata di LIM e non necessariamente in laboratorio. L'attività consiste nella costruzione di grafici orari attraverso "camminate" fatte dagli stessi studenti davanti al sensore di posizione. L'utilizzo dei sensori per misurare grandezze fisiche collegate al proprio corpo (per esempio la velocità nel caso del sensore di posizione) facilita la costruzione di conoscenza, come è stato mostrato in studi sviluppati nell'ambito della teoria cognitiva dell'*Embodied Mind Thesis* [14]; secondo questa teoria, l'attività cognitiva è strettamente correlata all'esperienza corporea e quindi se lo studente può sperimentare attraverso il proprio corpo ciò che accade nell'atto della misura e della rappresentazione grafica sarà agevolato nella costruzione dei concetti matematici necessari alla modellizzazione dei fenomeni. All'attività pratica segue l'analisi quantitativa dei grafici ottenuti, consentendo al docente di passare alla formalizzazione dei concetti introdotti [15].

**3.2 Dispositivi portatili multimediali**

I dispositivi portatili multimediali, *smartphone*, *tablet*, ecc. – dispositivi elettronici con funzionalità aggiuntive e personalizzabili a scelta dell'utilizzatore con l'installazione di *software* aggiuntivi (app) – sono sempre più diffusi tra i giovani. Effettuare esperimenti utilizzando strumenti che gli studenti posseggono e conoscono potrebbe stimolare il loro interesse per lo studio delle discipline scientifiche e, in particolare, della Matematica e della Fisica [16,17]. Nell'ambito del Laboratorio TIC, sono state proposte diverse attività laboratoriali basate sui DPM.

I DPM specialmente gli *smartphone* possono essere usati come strumenti di laboratorio, in quanto equipaggiati con un notevole numero di sensori, tra cui vi è l'accelerometro. A differenza del tachimetro di cui sono dotati i mezzi di trasporto, l'accelerometro è uno strumento poco conosciuto il cui principio di funzionamento non è di immediata comprensione. Un semplice accelerometro è costituito da una massa di prova, montata su delle molle, che può muoversi liberamente in una direzione, come illustrato in Figura 1(a) e 1(b). Un modello meccanico di tale accelerometro è stato proposto dagli autori [18] come strumento utilizzabile nella didattica per introdurre il principio di funzionamento tramite l'analisi del fenomeno fisico che permette di misurare accelerazioni. Conoscere il meccanismo su cui si basa uno strumento e poterlo facilmente riprodurre facilita la comprensione della relazione fra elemento sensibile e misura (a tale proposito, si pensi al valore didattico della costruzione di un termoscopio prima dell'utilizzo del termometro).

Quando il dispositivo su cui è installato l'accelerometro cambia velocità, la massa di prova si sposta dalla sua posizione di equilibrio. Questa variazione di posizione della massa può essere misurata con vari metodi. Nel modello meccanico si sfrutta l'allungamento delle molle, negli smartphone si sfrutta invece l'effetto capacitivo creato da tre lamine, di cui due laterali fisse e una centrale mobile, che formano due condensatori elettrici collegati in serie [19], come illustrato in Figura 1(c). Quando il dispositivo accelera, la lamina centrale mobile si sposta facendo cambiare la capacità elettrica del sistema. Dalla misura del valore della capacità si può determinare il valore dell'accelerazione [20]. La possibilità di disporre di questo sensore consente di utilizzare i DPM per una notevole varietà di esperimenti di fisica [21].

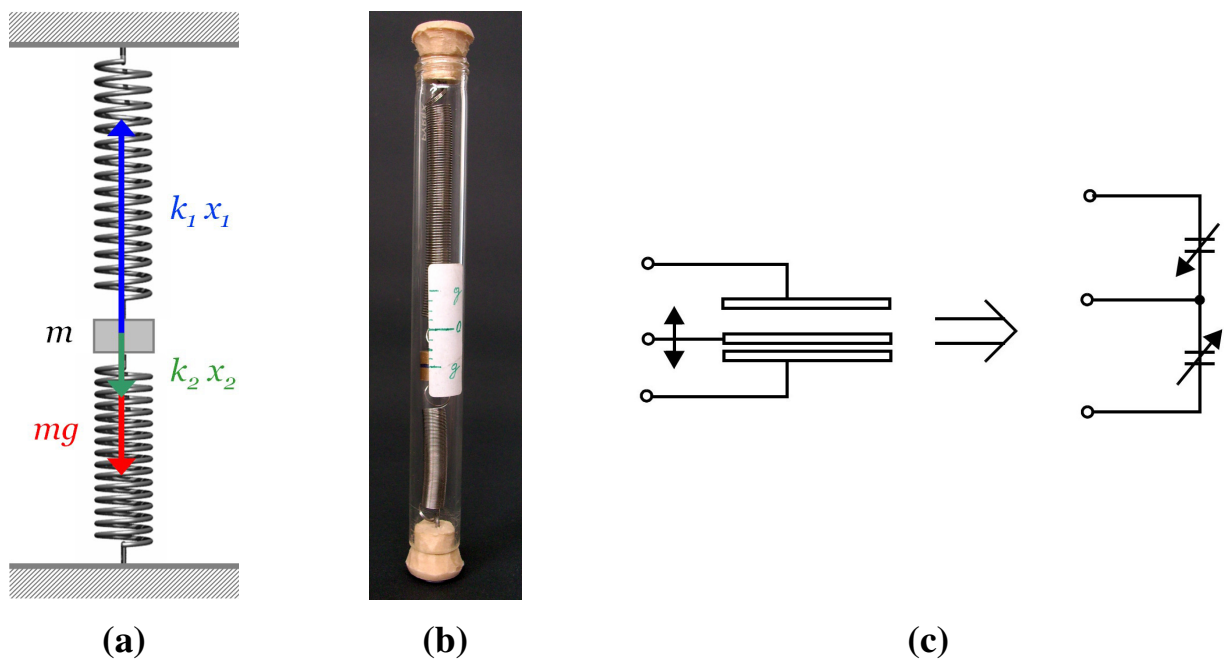

**Figura 1.** Schema del modello meccanico dell'accelerometro e di quello capacitivo a confronto; (a) modello meccanico [18]; (b) immagine dello strumento di Ref. [18]; (c) modello capacitivo e schema circuitale [19].

I sistemi operativi installati nei DPM consentono generalmente di leggere direttamente il valore dell'accelerazione, ma installando nel proprio *smartphone* specifiche applicazioni è possibile e riportare in grafico i valori rilevati in funzione del tempo, in questo modo si può per esempio realizzare il seguente esperimento. Registrando l'accelerazione dello *smartphone* durante la sua caduta libera sopra un morbido cuscino, si otterrà un grafico dell'accelerazione in funzione del tempo in cui l'accelerazione è zero durante la caduta libera e avrà un certo valore non nullo quando lo *smartphone* urta il cuscino.

Tra le applicazioni che usano l'accelerometro e che possono essere usate per la realizzazione di esperimenti di meccanica, vi è l'inclinometro (o livella a bolla). Questa applicazione consente di usare lo *smartphone* per la misura dell'inclinazione di un piano inclinato oppure delle componenti della forza di gravità nel piano verticale.

## 4. *Software* per l'insegnamento/apprendimento della Matematica e della Fisica

Un altro aspetto delle tecnologie informatiche nell'ambito della didattica della Matematica e della Fisica riguarda l'uso di *software* per la modellizzazione e per la simulazione. Tali *software* possono essere utilizzati in alternativa o a supporto delle attività in laboratorio; essi si rivelano particolarmente utili nelle non poche realtà scolastiche nelle quali non esiste un laboratorio di fisica o è difficoltoso frequentarlo.

### 4.1. Strumenti *open source* per la modellizzazione Matematica

Esistono oggi diversi ambienti di modellizzazione matematica. Nel Laboratorio TIC si è scelto di trattare il *software open source* Geogebra, un software di geometria dinamica, sviluppato da Judit e Marcus Hoenwarter [22] e coadiuvati da vari collaboratori. Geogebra è un ambiente aperto, pensato per l'utilizzo in qualsiasi livello scolastico e unisce dinamicamente geometria, algebra, tabelle, grafici, statistica, ecc; è molto usato in ambito matematico nella costruzione di modelli e consente un ampio utilizzo nella didattica della Fisica, risultando particolarmente adatto alla modellizzazione vettoriale e all'ottica geometrica. Qui di seguito riportiamo due esempi realizzati da studenti del Liceo Scientifico "Galileo Galilei" di Palermo. In Figura 2 è mostrata la scomposizione della forza peso di un corpo sopra un piano inclinato, mentre in Figura 3 è mostrata la modellizzazione geometrica del Mirascope, strumento basato sulla formazione di immagini virtuali nella doppia riflessione della luce in due specchi parabolici.

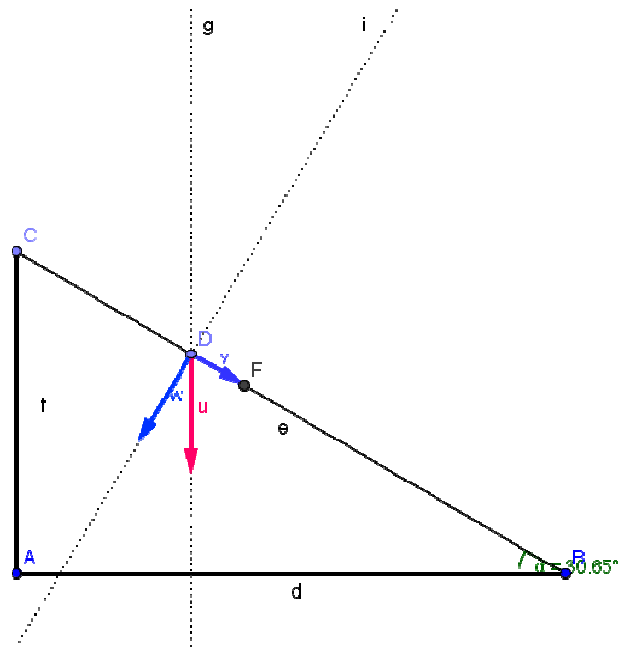

**Figura 2.** Scomposizione della forza peso di un corpo sopra un piano inclinato.

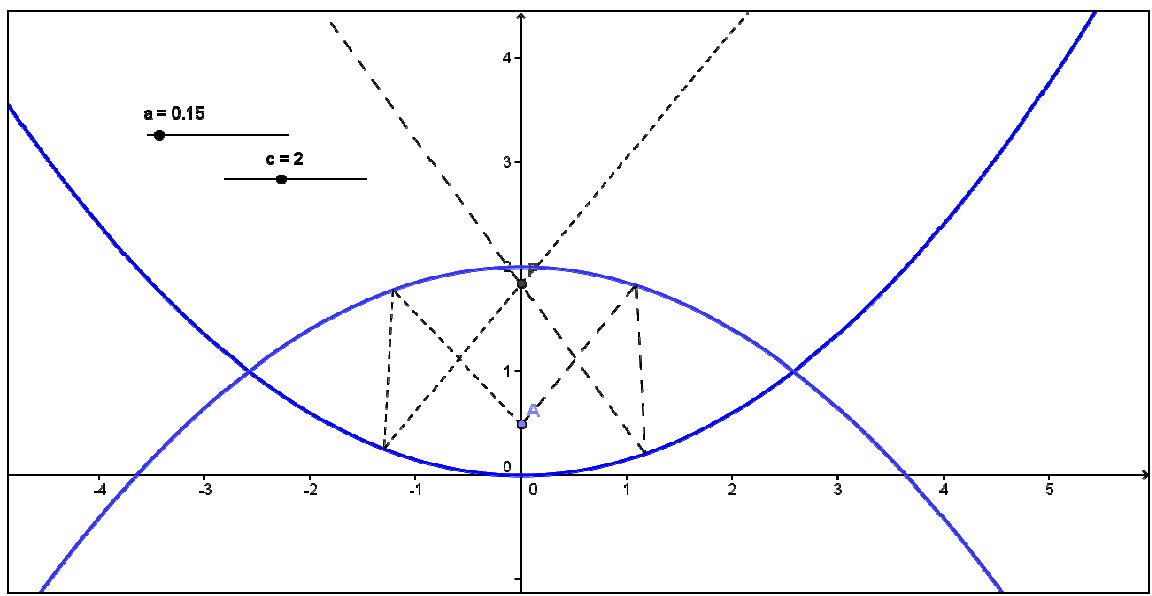

**Figura 3.** Modellizzazione geometrica del Mirascope, strumento basato sulla formazione di immagini virtuali nella doppia riflessione della luce in due specchi parabolici; **a** e **c** sono i coefficienti delle equazioni delle parabole e possono essere variati dagli *slider*.

Le sue potenzialità lo rendono adeguato a un ampio utilizzo laboratoriale, sia nella modalità di *problem solving* in cui viene proposto agli studenti un problema e viene richiesta la costruzione di un modello risolutivo, sia in modalità dimostrativa in cui il docente predispone un foglio di lavoro con il quale lo studente può interagire singolarmente o in gruppo. L'utilizzo di Geogebra per la modellizzazione matematica è stato trattato nei corsi di didattica e di laboratorio della Matematica, mentre nel Laboratorio TIC l'attività proposta ha riguardato solamente una riflessione critica sulla possibilità di usare questo strumento come parte integrante della formazione a distanza.

**4.2. La simulazione in Fisica**

Analogamente a quanto detto per la Matematica, sono disponibili in rete diversi *software* per la simulazione di fenomeni fisici e i relativi archivi di simulazioni [23, 24]. Nel corso del Laboratorio TIC è stato analizzato a titolo esemplificativo il sito dell'Università del Colorado [24], dove è possibile trovare varie simulazioni. Una simulazione che ha suscitato particolare interesse nel gruppo dei tirocinanti è stata quella in cui la sagoma dell'attore John Travolta produce delle scintille sulla maniglia di una porta dopo che la scarpa è stata strofinata su un tappeto [25], come illustrato in Figura 4. La simulazione, semplice e di sicuro effetto, permette di descrivere e modellizzare il noto fenomeno di vita quotidiana legato all'elettricità statica. L'analisi di questa simulazione ha permesso una riflessione sulla potenza didattica della modellizzazione dei fenomeni fisici a partire dall'esperienza quotidiana e sull'importanza della presentazione di problemi e modelli in una modalità ludica.

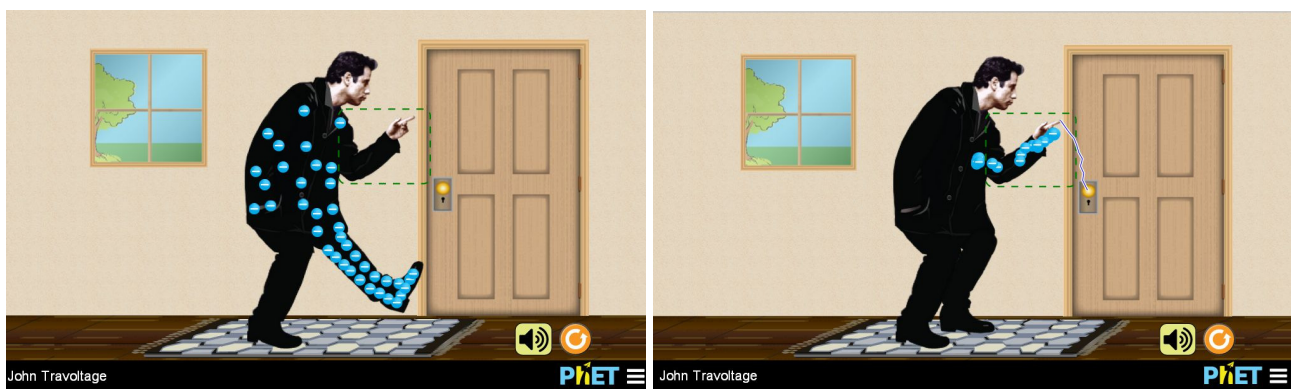

**Figura 4.** Due momenti della simulazione "John Travoltage": il corpo si carica per strofinio della scarpa e successivamente si scarica quando si avvicina alla maniglia della porta [25].

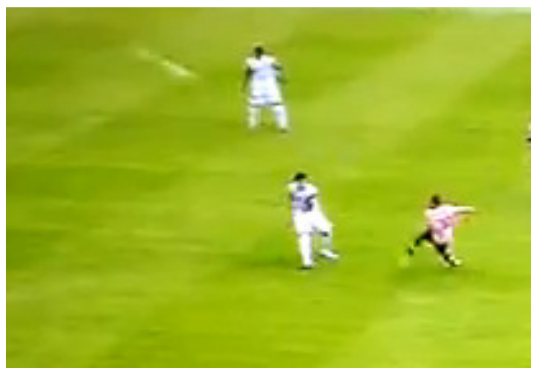

**Figura 5.** Il fotogramma, tratto da un filmato, permette di intuire la traiettoria del pallone calciato dal giocatore sulla destra.

Durante il corso è stato inoltre presentato il micro-mondo newtoniano *Interactive Physics* [23] con una dimostrazione pratica di come sia possibile, attraverso l'utilizzo dell'ambiente aperto, costruire in classe situazioni di apprendimento che sfruttano il ciclo previsione-esperimento-confronto (PEC) [26]. Nello specifico, è stato analizzato il caso del moto di un grave, evidenziando come sia molto semplice costruire la simulazione in maniera cooperativa utilizzando la LIM e seguendo i passi della modellizzazione [27] per poi sviluppare il ciclo PEC. L'esempio trattato parte dall'osservazione della traiettoria del pallone in una partita di calcio (Figura 5), fenomeno noto e rappresentativo del ben più ampio fenomeno del "moto del proiettile". In maniera cooperativa e pratica, si individuano le grandezze utili alla descrizione del fenomeno (velocità, in modulo e direzione), si descrive il fenomeno (in questo è utile la visualizzazione stroboscopica) e si determinano le relazioni fra le variabili utilizzando grafici qualitativi (velocità maggiore, maggiore altezza raggiunta, a parità di inclinazione…), per poi passare alla formalizzazione matematica (Figura 6).

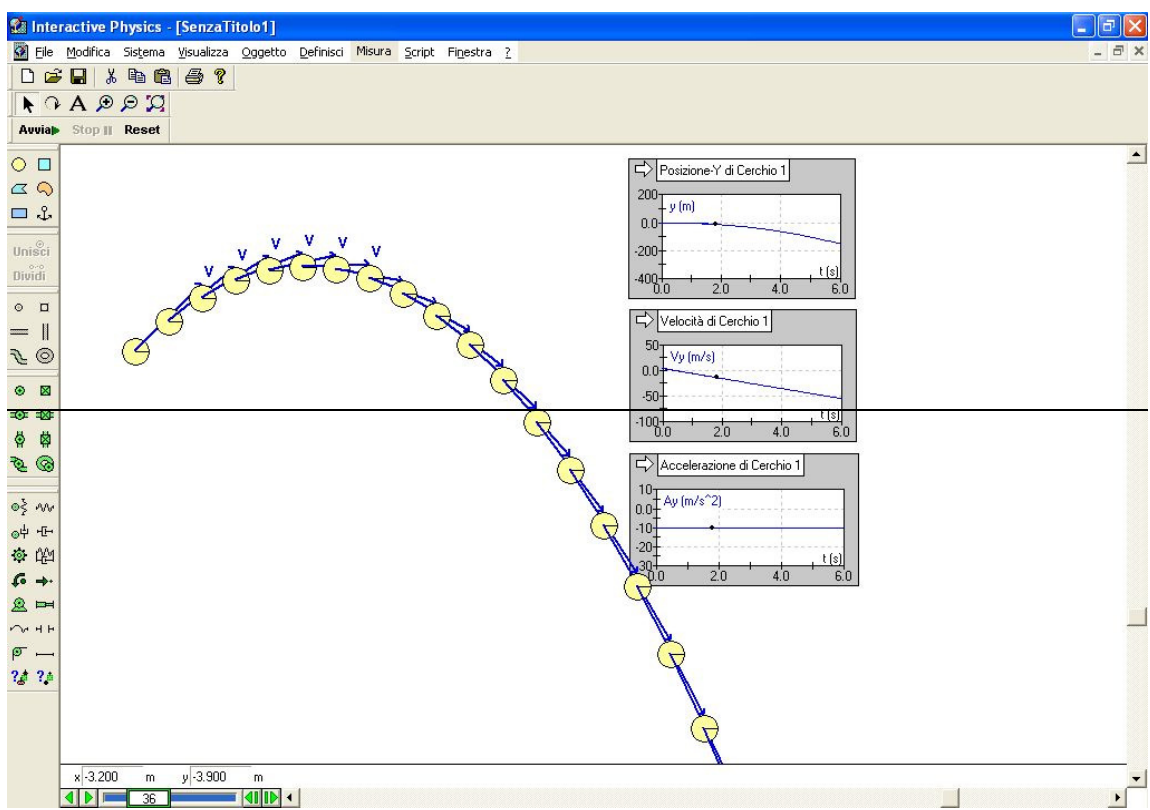

**Figura 6.** Traccia stroboscopica del moto di un corpo lanciato verso l'alto con una velocità inclinata rispetto all'orizzontale, i grafici mostrano le leggi orarie della componente verticale del moto.

### 4.3. Analisi video: il *software* Tracker

Tracker è un *software* libero sviluppato nell'ambito del progetto *Open Source Physics* [28,29]. Esso permette di effettuare l'analisi al computer di filmati di moti di oggetti [30,31]. Proprio per questa sua caratteristica, questo *software* può essere implementato nelle attività di laboratorio di meccanica al liceo e in particolare per lo studio della cinematica. Infatti, con questo tipo di attività possono essere realizzati esperimenti efficaci e a basso costo, con un *setting* che prevede oggetti di uso quotidiano di cui studiare la meccanica del moto, una riga graduata, uno *smartphone* per le riprese e il *software* Tracker per l'analisi dei video realizzati. Tale strumento risulta essere di grande interesse nella didattica della meccanica sia di primo biennio (specialmente meccanica unidimensionale) sia di terzo anno (specialmente dinamica nel piano e urti). In Figura 7 è mostrata l'analisi del moto in caduta libera di un oggetto effettuata con Tracker, da cui si ottengono i grafici di Figura 7(b) che descrivono chiaramente l'andamento temporale dello spostamento, della velocità e dell'accelerazione nel moto rettilineo uniformemente accelerato.

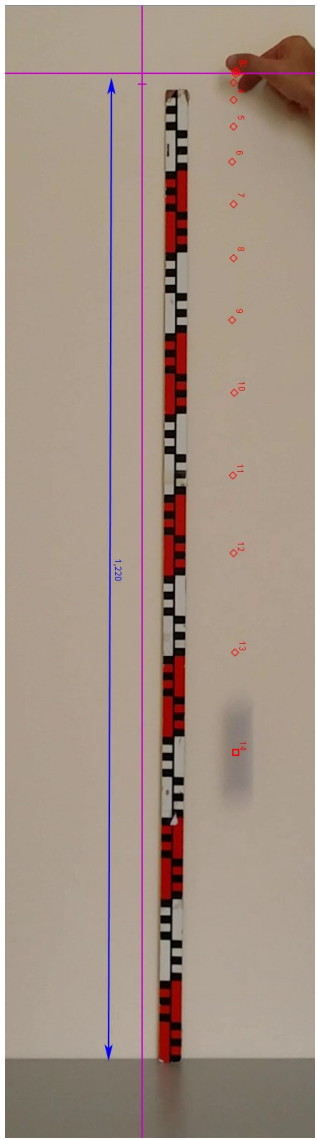
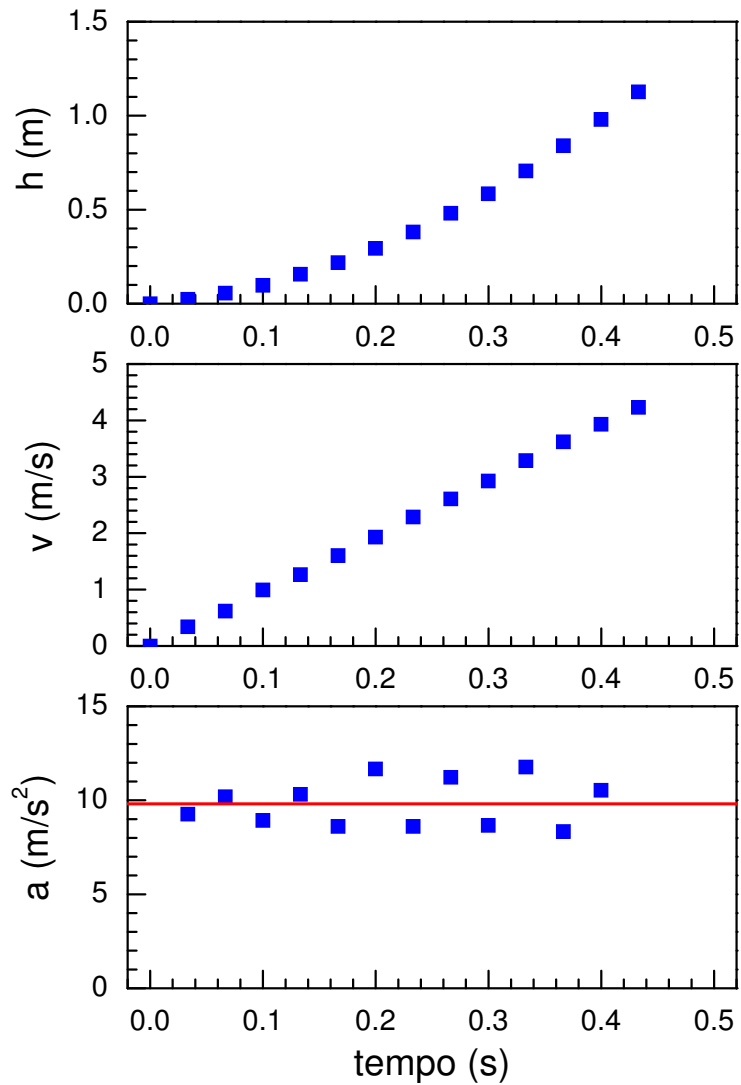

**(a)** **(b)**

**Figura 7.** Moto in caduta libera (a) e relativo grafico della posizione, velocità e accelerazione in funzione del tempo (b) ottenuti con i dati acquisiti; l'origine degli assi è fissato nel punto a quota più alta; la linea rossa nel grafico dell'accelerazione indica il valore di 9.81 m/s$^2$.

## 5. Discussione e conclusione

Il Laboratorio TIC ha permesso ai tirocinanti di riflettere sulla varietà di strumenti tecnologici/informatici e comunicativi che possono essere utilizzati nel lavoro quotidiano a Scuola in ambito didattico. Un docente ha il compito di favorire la costruzione di conoscenza negli studenti educandoli a essere cittadini consapevoli e responsabili e ciò non può avvenire se il docente non è formato e aggiornato continuamente sulle innovazioni tecnologiche soprattutto nella didattica. Una formazione iniziale sul modello di quella messa in atto nel TFA (precedentemente nelle SIS) a nostro avviso mette il laureato, che vuole intraprendere la professione di docente di Scuola Secondaria, di fronte alla necessità di conoscere dall'interno il mondo della Scuola. Nel contempo, l'interazione con l'Università offre la possibilità di conoscere il mondo della ricerca, dove si sviluppano gli strumenti tecnologici utilizzabili in ambito scolastico/didattico. Si innesca così il meccanismo virtuoso della conoscenza che crea e richiede altra conoscenza. Il docente di nuova generazione, formato in un ambiente ricco di stimoli come può essere quello in cui si stabilizza la collaborazione osmotica fra Università e Scuola, potrà in questo modo crescere professionalmente.

In conclusione, nell'articolo è stata presentata l'organizzazione del Laboratorio Pedagogico Didattico TIC finalizzata da un lato alle novità nella gestione amministrativa/burocratica della Scuola e dall'altro allo sviluppo delle tecnologie dell'informazione e della comunicazione e la loro applicazione in ambito didattico. Sono state discusse le varie possibilità di rinnovamento della didattica introdotte dagli strumenti di comunicazione e condivisione del *web* 2.0, dall'uso della LIM, fino all'uso dei DPM come strumenti di laboratorio.

## Ringraziamenti e note biografiche

**Lucia Lupo** è docente in ruolo presso il Liceo Scientifico Statale "Galileo Galilei" di Palermo, ha tenuto l'insegnamento di Laboratorio Pedagogico Didattico TIC e ha seguito gli studenti tirocinanti nel loro percorso di tirocinio in qualità di Tutor Coordinatore della Classe A049 Matematica e Fisica.

**Aurelio Agliolo Gallitto** è professore associato nel SSD FIS/01 ed è referente del MIUR per il TFA Classe 049 Matematica e Fisica presso l'Università di Palermo.

## Bibliografia, sitografia e riferimenti normativi